\documentclass[onecolumn,twoside,slac_two]{revtex4}
\usepackage{graphicx,amsmath}
\usepackage{fancyhdr}
\begin{document}

\title{ $B$ decays into a scalar/tensor meson in pursuit of determining the CKM angle $\gamma$}

%

\author{Wei Wang}
\affiliation{Deutsches Elektronen-Synchrotron DESY, D-22607 Hamburg, Germany}

\begin{abstract}
In this work, I suggest a new way for determining the CKM angle $\gamma$ via $B$ decays into a scalar/tensor meson  without any hadronic uncertainty. The proposed idea makes profits of the two triangles formed by the  $B^\pm\to (D^0,\bar D^0,D_{CP}^0) K^{*\pm}_{0(2)}(1430)$ decay amplitudes.  The advantages in it  are  large CP asymmetries and the avoidance of the use of doubly Cabibbo-suppressed $D$ decays. Branching ratios of  $B^\pm\to (D^0,\bar D^0,D_{CP}^0) K^{*\pm}_{0(2)}(1430)$ are estimated to have the order $10^{-6}-10^{-5}$ and therefore measurable by the ongoing LHCb experiment and future  experimental facilities. 
The usefulness of other related channels, for instance the neutral $B_d$ decays into $DK^{*}_{0(2)}(1430)$ and   $B_s\to (D^0, \bar D^0) M$ ($M=f_0(980),f_0(1370),  f_2'(1525), f_1(1285), f_1(1420), h_1(1180))$, the  $B\to D^{\mp} a_{0,2}^{\pm}$  for the extraction of $\gamma+2\beta$ and the $B_s\to D^{\mp} K_{0,2}^{*\pm}$ to access $\gamma+2\beta_s$,  is also discussed in brevity. 

\end{abstract}

\maketitle

\thispagestyle{fancy}


\section{INTRODUCTION} 

The standard model description of CP violation is encoded in a single, irreducible phase in the $3\times 3$ quark mixing matrix called the Cabibbo-Kobayashi-Maskawa (CKM) matrix.   One of the foremost tasks in the past decades has been to 
study the   $(bd)$ unitarity triangle, the graphical
representation of the condition stemming from  the unitarity
of the CKM matrix: $V_{ud}V_{ub}^*+V_{cd}V_{cb}^*+V_{td}V_{tb}^*=0$, whose  angles  $(\alpha,\beta,\gamma)$ satisfy $\alpha+\beta+\gamma=180^\circ$.  
In contrast with the precise results on $\alpha$ and $\beta$~\cite{Asner:2010qj,CKMfitter} , our knowledge of  the angle $\gamma$ is rather uncertain, with a precision of roughly $10^\circ$. This is one of  the dominant sources of the current uncertainties on the apex of the  $(bd)$ unitary triangle. 
One of the major efforts in flavor physics by LHCb and  the future SuperB factory experiments will be devoted to reduce the errors in $\gamma$.

 In this work~\cite{Wang:2011zw}, we propose a new method  based on $B\to D M$ decays with $M$ being a light scalar/tensor meson. Among the various   modes to be discussed, of special interest  are the $B^\pm\to (D^0,\bar D^0,D_{CP}^0) K^{*\pm}_{0(2)}(1430)$ modes,  where $K^*_{0(2)}(1430)$ is a $J^P=0^+(2^+)$ scalar (tensor) resonance. The small (zero) decay constant of $K^*_{0}(1430)$($K^*_{2}(1430)$) compensates the large Wilson coefficient in  the color-allowed amplitude, leading to comparable decay amplitudes of $B^\pm\to D^0 K^{*\pm}_{0(2)}(1430)$ and $B^\pm\to\bar D^0 K^{*\pm}_{0(2)}(1430)$. As a consequence,   large CP asymmetries are predicted. Measurements of  branching ratios (BRs) of the neutral $B_d$ decays into $DK^{*}_{0(2)}(1430)$ and  time-dependent CP asymmetries in $B_s\to D M$ ($M=f_0(980),f_0(1370),  $ $ f_2'(1525), f_1(1285), f_1(1420), h_1(1180))$ are also useful to extract the angle $\gamma$.  In addition, the combination $\gamma+2\beta$ and $\gamma+2\beta_s$ could be extracted  via time-dependent measurements of the $B\to D^\pm a_{0,2}^\mp$  and  the $B_s\to D_s^\pm K_{0,2}^{*\mp}$ which  may   have large CP asymmetries.

\section{$\gamma$ from $B \to D^0(\bar D^0) K^{*}_{0,2}(1430)$  }\label{sec:K*0}

The angle $\gamma\equiv arg(- V_{ud}V_{ub}^*/(V_{cd}V_{cb}^*))$ is the relative weak phase of the decays  induced by the $b\to c\bar us$ and $b\to u\bar cs$ transition. 
The proposed method in this work uses the information that the six decay amplitudes of $B^\pm \to (D^0, \bar D^0, D_{CP}^0)K_{0,2}^{*\pm}$  form two triangles in the complex plane,  graphically manifesting the  identities
\begin{eqnarray}
 \sqrt 2 A(B^+\to D_\pm^0 K^{*+}_{0,2}) = A(B^+ \to D^0K^{*+}_{0,2})  \pm A(B^+\to \bar D^0K^{*+}_{0,2}),\nonumber\\
 \sqrt 2 A(B^-\to D_\pm^0 K^{*-}_{0,2}) = A(B^- \to D^0K^{*-}_{0,2})   \pm A(B^-\to \bar D^0K^{*-}_{0,2}),\label{eq:identity}
\end{eqnarray} 
in which we have adopted the convention $CP|D^0\rangle =|\bar D^0\rangle$  and $D^0_\pm =(|D^0\rangle \pm|\bar D^0\rangle)/\sqrt2 $ are the CP even and odd eigenstates.    
Measurements of the decay rates of the six processes have the potential to completely determine the sides and apexes of the two triangles,  
more particularly the relative phase between $A(B^- \to \bar D^0K^{*-}_{0,2})$ and its CP conjugate $A(B^+ \to D^0K^{*+}_{0,2})$ is $2\gamma$. 

The ratio of the sides  and the relative strong phase difference govern
the shape of the two triangles
\begin{eqnarray}
 r_{B}^{K_J}\equiv\left|{A(B^-\to \bar D^0 {K_J^-})}/{A(B^-\to D^0 K^{-}_J)}\right|, 
 \delta_{B}^{K_J} \equiv arg\left[{e^{i\gamma} A(B^-\to \bar D^0 K^{-}_J)}/{A(B^-\to D^0 K^{-}_J)}\right],\nonumber
\end{eqnarray}
with $K_J= K^*_{0,2}$. In fact  physical observables to be experimentally measured are   defined in terms of these two quantities:
\begin{eqnarray}
 R_{CP\pm}^{K_J} &=&2\frac{{\cal B}(B^-\to D_{CP\pm} K_J^-)+{\cal B}(B^+\to D_{CP\pm} K_J^+)  }{{\cal B}(B^-\to D^0K_J^-) +{\cal B}(B^+\to \bar D^0 K_J^+) } 
= 1+(r_{B}^{K_J})^2\pm 2r_{B}^{K_J} \cos\delta_{B}^{K_J} \cos\gamma,\nonumber\\
 A_{CP\pm}^{K_J} &=&\frac{{\cal B}(B^-\to D_{CP\pm} K_J^-)-{\cal B}(B^+\to D_{CP\pm} K_J^+)  }{{\cal B}(B^-\to D_{CP\pm} K_J^-) +{\cal B}(B^+\to D_{CP\pm} K_J^+) } =\pm 2r_B^{K_J} \sin\delta_{B}^{K_J} \sin\gamma /R_{CP\pm}^{K_J}.  \nonumber 
 \label{eq:experiments}
\end{eqnarray}

Up to now,   experimental results on $\gamma$ mostly come from the $B^\pm\to  D {K^\pm }$~\cite{CKMfitter,Aaij:2012kz}, in which the  $B^-\to \bar D^0 K^-$ is suppressed by  both CKM factor and color factor.  The ratio of decay amplitudes $r_B^K \sim |V_{ub}V_{cs}^*/(V_{cb}V_{us}^*) a_2/a_1|\sim 0.1$ is small and thereby the two triangles formed by decay amplitudes are very squashed.
$ R_{CP\pm}^K $ and $ A_{CP\pm}^K$ have a mild sensitivity to the angle $\gamma$, inducing large experimental uncertainties in   $\gamma$~\cite{CKMfitter,Aaij:2012kz}. 

Here we wish to stress that  the low sensitivity problem is highly improved  in $B^\pm\to  D {K^{*\pm}_{0,2} }$ due to $r_{K_{0,2}^*}\sim 1$ and in particular large CP asymmetries are expected. Although the color-allowed diagram has a large Wilson coefficient $a_1\sim 1$,  the emitted ${K_{0,2}^*}$ meson is produced from a local vector or  axial-vector current (at the lowest order in $\alpha_s$), whose matrix element between the QCD vacuum and  the $K^*_{0}$($K^*_{2}$) state is small (identically zero).  Due to the suppression from decay constant, the color-allowed amplitude is comparable to the color-suppressed one giving $r_{K_{0,2}^*}\sim 1$.

\begin{figure}\begin{center}
\includegraphics[scale=1.0]{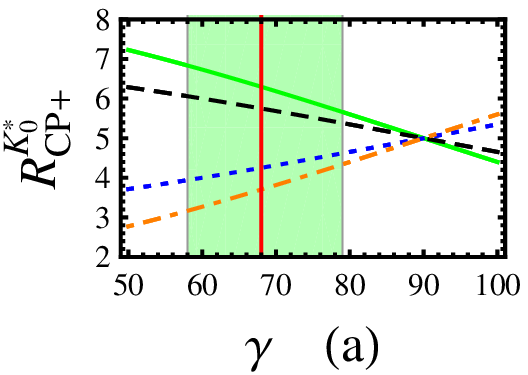}
\includegraphics[scale=1.05]{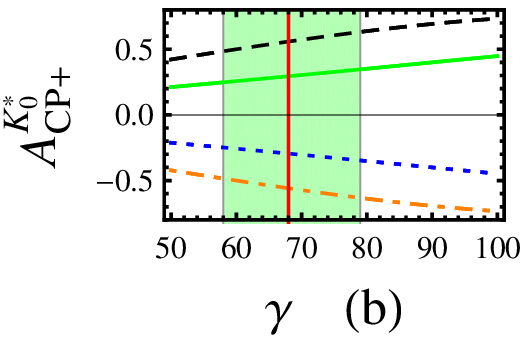}
\includegraphics[scale=1.0]{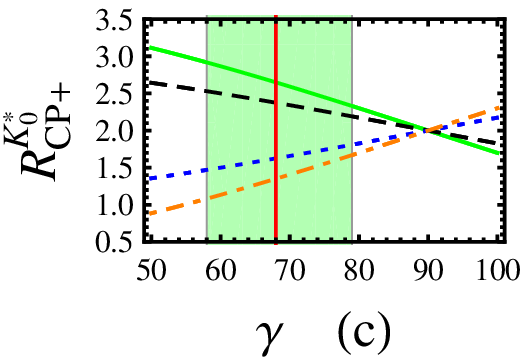}
\includegraphics[scale=1.05]{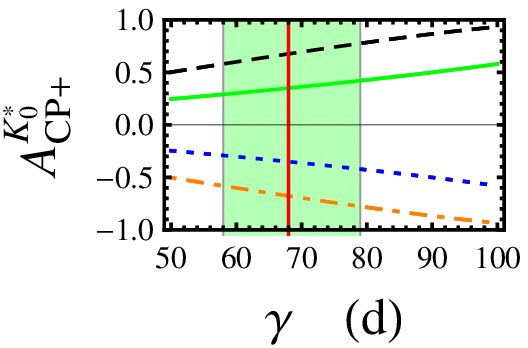}
\includegraphics[scale=1.0]{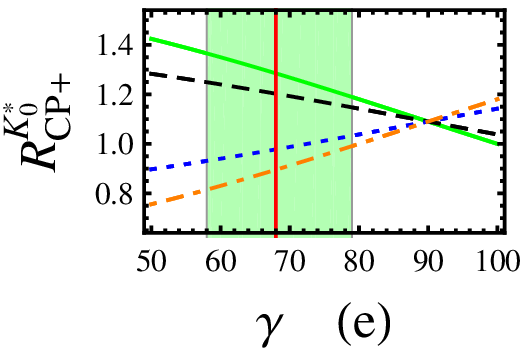}
\includegraphics[scale=1.05]{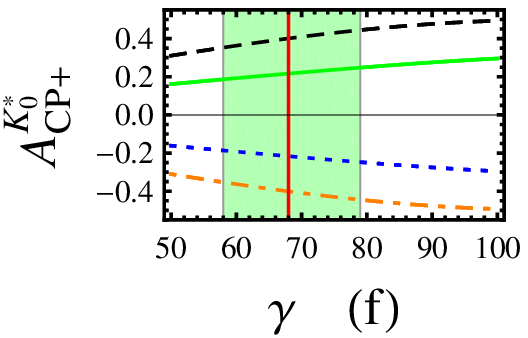}
\caption{ The sensitivity of   $R_{CP+}^{K^*_0}$ and $A_{CP+}^{K^*_0}$ to $\gamma$. In panels (a,b), $r_{B}^{K^*_0}=2$ is employed, while in panels (c,d) $r_{B}^{K^*_0}=1$. In the last two panels (e,f), we consider the case in which the ratio is not enhanced sizably $r_{B}^{K^*_0}=0.3$. Such choice  corresponds to the sign of  the Wilson coefficient $a_2$ reversed namely $a_2=-0.2$,  resulting in a large branching ratio: ${\cal B}(B^-\to D^0 K^{*-}_{0,2})\sim10^{-5}$. 
In this figure,  the solid (green), dashed (black), dotted (blue) and dot-dashed (orange) lines in diagrams (a,c,e) are obtained according to four choices $\delta_{B}^{K^*_0}=(30,60,120,150)^\circ$ respectively,   while the corresponding curves in panels (b,d,f) correspond to $\delta_{B}^{K^*_0}=(30,60,-30,-60)^\circ$. The shadowed  (light-green) region denotes the current bounds on $\gamma=(68^{+10}_{-11})^\circ$ from a combined analysis of $B^\pm\to DK^\pm$~\cite{CKMfitter}, in which the vertical (red) line reflects the central value.   
 } \label{fig:dependence}
\end{center}
\end{figure}

An estimate of  decay amplitudes can be made  under the factorization assumption
\begin{eqnarray}
&& A(B^-\to \bar D^0 K^{*-}_0) =-V_{ub}V_{cs}^*  C,  
 \;\; A(B^-\to  D^0 K^{*-}_0) = -  V_{cb}V_{us}^* (C-T),\label{eq:factorization}
\end{eqnarray}
where 
\begin{eqnarray}
C= G_F f_{D} a_2(m_{B}^2-m_{K^*_0}^2)  F_0^{B K^*_0}(m_D^2)/\sqrt 2,\;\;
T=G_F f_{K^*_0}a_1(m_{B}^2-m_{D}^2)  F_0^{BD}(m_{K^*_0}^2)/\sqrt 2, \nonumber
\end{eqnarray}
 and
$G_F$ is the Fermi constant.  The $K^*_0$ decay constant vanishes in the flavor SU(3) symmetry limit and may acquire a small value arising from the symmetry breaking effects. Adopting the results in QCD sum rules~\cite{Cheng:2005nb}
\begin{eqnarray}
 f_{K^*_0}=-24 {\rm MeV}, \;\;\; {\rm or }  \;\;\;
 f_{K^*_0}=36 {\rm MeV}, \nonumber
\end{eqnarray}
we obtain the relation $2a_1  |f_{K^*_0}|\sim a_2 f_{D}$.
Using the $B\to K^*_0$ form factors computed in the perturbative QCD approach~\cite{Li:2008tk}, the $B\to D$ form factors from Ref.~\cite{Cheng:2003sm} and $a_2=0.2, a_1=1$ 
we get  $C/T\sim 1.2$  and
\begin{eqnarray}
r_B^{K^*_0}=\left|C{V_{ub} V_{cs}^*}/[{V_{cb}V_{us}^*}(C-T)] \right|\sim 2,\;\;\;
{\cal B}(B^-\to \bar D^0 K^{*-}_0)
\sim 
4\times 10^{-6}.
\end{eqnarray}

Since the strong phase can not be computed by perturbation theory, we shall take several benchmark values to illustrate the dependence of $R_{CP+}^{K^*_0}$  and $A_{CP+}^{K^*_0}$ in Fig.~\ref{fig:dependence}. In panels (a,b), $r_{B}^{K^*_0}=2$ is employed, while in panels (c,d) $r_{B}^{K^*_0}=1$. In the last two panels (e,f), we consider the case in which the ratio is not enhanced sizably $r_{B}^{K^*_0}=0.3$. Such choice  corresponds to the sign of  the Wilson coefficient $a_2$ reversed namely $a_2=-0.2$,  resulting in a large branching ratio: ${\cal B}(B^-\to D^0 K^{*-}_{0,2})\sim10^{-5}$. 
In this figure,  the solid (green), dashed (black), dotted (blue) and dot-dashed (orange) lines in diagrams (a,c,e) are obtained according to four choices $\delta_{B}^{K^*_0}=(30,60,120,150)^\circ$ respectively,   while the corresponding curves in panels (b,d,f) correspond to $\delta_{B}^{K^*_0}=(30,60,-30,-60)^\circ$. The shadowed  (light-green) region denotes the current bounds on $\gamma=(68^{+10}_{-11})^\circ$ from a combined analysis of $B^\pm\to DK^\pm$~\cite{CKMfitter}, in which the vertical (red) line reflects the central value.    CP odd quantities can be obtained  via the relation in Eq.~\eqref{eq:experiments}, for instance     $R_{CP-}^{K^*_0}=(R_{CP+}^{K^*_0})_{\delta_{B}^{K^*_0}\to 180^\circ - \delta_{B}^{K^*_0}}$.

\section{$\gamma$ from other useful channels}

The above method to use the two triangles formed by the six decay amplitudes for determining $\gamma$ is also valid in    $B_d\to DK_{0,2}^{*0}$, in which the tree amplitude $T$ vanishes.  No time-dependent measurement is required since $K^{*}_{0,2}$ is self-tagging.  The color-suppressed decay amplitudes involving $D^0$ and $\bar D^0$ arise from the same type of Feynman diagram, thereby one may expect that $\delta_{B}^{K^*_0}\sim 0$. If true,  the CP asymmetries $A_{CP\pm}^{K^*_0}$ would be roughly 0 but $R_{CP\pm}^{K^*_0}$ can sizably deviate from 1.

Following the similar strategy, we collect in Tab.~\ref{Tab:gammaangles} other  $B$ decay channels  into a scalar/tensor meson useful for the extraction of CKM angle $\gamma$ or its combination. All these decay modes are expected to have larger $r_f$, ratios of decay amplitudes, compared to the corresponding channels in which the scalar/tensor meson is replaced by a pseudoscalar pion/kaon.  For earlier discussions  on $B$ decays into a scalar/tensor meson and the role in extracting $\gamma+2\beta$, please see Ref.~\cite{Diehl:2001xe}. The branching fractions and ratios of decay amplitudes of the $B\to DT$ modes are taken from the recent calculation in the perturbative QCD approach~\cite{Zou:2012sx} while the rest  entries  when available are obtained in the factorization approximation~\cite{Wang:2011zw} with   inputs from Refs.~\cite{Li:2008tk,Wang:2010ni,Colangelo:2010bg}.   Since  the branching ratios obtained in naive factorization method are very small and usually suffer large theoretical uncertainties, an analysis in which the QCD/power corrections are taken into account is called for.

To illustrate how to extract the CKM angle, we consider  the example  of $B_s\to Df_0$ with the amplitudes
\begin{eqnarray}
A(\bar B_s\to \bar  D^0f_0) = V_{ub}V_{cs}^* A_1, 
A(B_s\to D^0f_0) = V_{ub}^*V_{cs}A_1,\nonumber\\
A(\bar B_s\to D^0f_0) = V_{cb}V_{us}^* A_2,
 A( B_s\to \bar D^0f_0) = V_{cb}^*V_{us} A_2.\label{eq:Bsdecayamplitudes}
\end{eqnarray}
For each amplitude, there is only one weak phase, and therefore no direct CP asymmetry is predicted.
For simplicity we use the notation for  the relative size and strong phase of the two amplitudes 
\begin{eqnarray}
 r_{B_s}^{f_0} = \left|{ V_{ub}V_{cs}^* A_1}/({V_{cb}V_{us}^* A_2})\right|,\;\;\; \delta_{B_s}^{f_0} = arg \left({A_1}/{A_2}\right).\label{eq:Bsratios}
\end{eqnarray}

The normalized time-dependent decay widths of $B_s\to Df_0$ are given by~\cite{Gronau:1989zb}
\begin{eqnarray}
 &&\Gamma(\bar B_s^0(t) \to D^0(\bar D^0) f_0)= e^{ -t/ \tau_{B_s}} \bar \Gamma \Big[1   + \cos(\Delta m t)C_{D^0(\bar D^0) f_0} 
+ \sin(\Delta mt)S_{D^0(\bar D^0) f_0}  \Big],
\end{eqnarray}
where $\bar \Gamma$  is the averaged decay width and we have neglected the width difference in the evolution. 
For the corresponding $B_s^0$ decays, the   signs in front of cosine and sine terms should   be reversed.   
Substituting the amplitudes defined in Eq.~\eqref{eq:Bsdecayamplitudes}, we have 
\begin{eqnarray}
 C_{D^0 f_0}  =  -C_{\bar D^0 f_0}  = [{1-(r_{B_s}^{f_0})^2}]/[{1+(r_{B_s}^{f_0})^2}],\nonumber\\
 S_{D^0 f_0}  = {-2 r_{B_s}^{f_0}  \sin( \gamma+ \delta_{B_s}^{f_0}+2\beta_s)}/[{1+(r_{B_s}^{f_0})^2}],\nonumber\\ 
 S_{\bar D^0 f_0}  =  {-2 r_{B_s}^{f_0}  \sin( \gamma- \delta_{B_s}^{f_0}+2\beta_s)}/[{1+(r_{B_s}^{f_0})^2}],
\end{eqnarray}
with the phase $\beta_s$   being the $B_s-\bar B_s$ mixing  phase $
 {q}/{p} = {V_{tb}^* V_{ts}}/({ V_{tb}V_{ts}^*}) 
= e^{-2i\beta_s}$. In the SM,  $\beta_s\simeq -0.019$ rad. 
The equality $C_{D^0 f_0} =- C_{\bar D^0 f_0}$ is  due to the fact that there is one  weak phase in decay amplitudes.  
Measuring the time-dependent decay widths, we will be able to determine the coefficients  $C_{D^0 (\bar D^0) f_0} $ and $S_{D^0 (\bar D^0) f_0} $, and thus the three quantities,  $r_{B_s}^{f_0}$,  $\delta_{B_s}^{f_0}$ and $\gamma+2\beta_s$, can be extracted cleanly modulo a discrete ambiguity.

 \begin{table}[htdp]
\caption{Properties of useful $B$ decay channels into a scalar/tensor meson towards the extraction of the CKM angle $\gamma$. 
The branching ratios and ratios of decay amplitudes of $B\to DT$ are taken from the recent calculation in the perturbative QCD approach~\cite{Zou:2012sx} while the rest  entries  when available are obtained in the factorization approximation~\cite{Wang:2011zw}.     }
\begin{center}
\begin{tabular}{|c|c|c|c|}\hline
 Channel & CKM angle to access & BRs for suppressed and allowed modes  & $r_f$ \\ \hline
 $B^\pm \to D^\pm K_{0}^*$ & $\gamma$  & [$4\times 10^{-6}$,  $4\times 10^{-5}$] & 0.3 \\ 
 $B^\pm \to D^\pm K_{2}^*$ & $\gamma$  & [$3\times 10^{-6}$,  $3\times 10^{-5}$] & 0.3  \\\hline
 $B\to D^\pm a_{0}^{\mp} $    & $\gamma+2\beta$ &   &   \\
 $B\to D^\pm a_{2}^{\mp} $    & $\gamma+2\beta$ & $[2\times 10^{-6}, 4\times 10^{-4}]$ &  0.1 \\\hline
 $B_s\to D_s^\pm K_{0}^{*\mp} $    & $\gamma+2\beta_s$ &   &    \\
 $B_s\to D_s^\pm K_{2}^{*\mp} $    & $\gamma+2\beta_s$ &$ [2\times 10^{-5},2\times 10^{-5}]$ &  1 \\\hline
 $B_s\to D f_0(980) $    & $\gamma+2\beta_s$ &$[1\times 10^{-6}, 3\times 10^{-6}]$ &  0.5 \\
 $B_s\to D f_2'(1525)$    & $\gamma+2\beta_s$ & $[3\times 10^{-6}, 1.4\times 10^{-5}]$ &  0.5\\\hline
 \end{tabular}
\end{center}
\label{Tab:gammaangles}
\end{table}%

 \section{Conclusions}
 
 Much progress has been made in recent years in testing the CKM description of the quark mixing and the CP violation. 
In this work we have explored the possibility to extract the CP violation angle $\gamma$ or the combination $\gamma+2\beta(\beta_s)$ with $B$ decays into a scalar/tensor meson.  A clean method is to use the two triangles formed by the decay amplitudes of  $B^\pm\to (D^0,\bar D^0,D_{CP}^0) K^{*\pm}_{0(2)}(1430)$.
We expect that $B^\pm\to D^0 K^{*\pm}_{0(2)}(1430)$ and $B^\pm\to\bar D^0 K^{*\pm}_{0(2)}(1430)$ have similar decay rates and the  CP asymmetries have a strong correlation with $\gamma$. The analysis is also supported by the recent study of $B\to DT$ in perturbative QCD approach.   Our method does not require the reconstruction   of the $D$  meson via its  doubly  Cabibbo-suppressed decays, which are usually buried under the combinatorial background.  Using the factorization assumption and the relevant experimental data we have  estimated   he branching ratios of these modes and find them  to be of order $10^{-5}- 10^{-6}$.   The usefulness of other related channels, for instance the neutral $B_d$ decays into $DK^{*}_{0(2)}(1430)$ and  the time-dependent CP asymmetries in  $B_s\to (D^0, \bar D^0) M$ ($M=f_0(980),f_0(1370),  f_2'(1525), f_1(1285), f_1(1420), h_1(1180))$, the  $B\to D^{\mp} a_{0,2}^{\pm}$  for the extraction of $\gamma+2\beta$ and $B_s\to D^{\mp} K_{0,2}^{*\pm}$ to access $\gamma+2\beta_s$,  is also discussed in brevity.

 \section*{Acknowledgement}
  I thank A.Ali and G.Hiller for useful discussions.
 This work is supported by Alexander von Humboldt Foundation.


\end{document}